\documentclass{iopart}

\usepackage[dvips]{graphicx}

 \expandafter\let\csname equation*\endcsname\relax
 \expandafter\let\csname endequation*\endcsname\relax

\usepackage{amsmath}

\begin{document}

\title[Efficient MER modeling]{Efficient Micro-electrode Recording Modeling using a Filtered Point Process}

\author{Kristian J. Weegink$^1$, Paul A. Bellette, John J. Varghese$^1$, Peter A. Silburn$^{2,3}$, Paul A. Meehan$^1$,  Andrew P. Bradley$^4$}

\address{$^1$University of Queensland, School of Mechanical and Mining Engineering, St Lucia, Queensland, Australia\\
$^2$University of Queensland, Asia-Pacific Center for Neuron Modulation, Herston, Queensland, Australia\\
$^3$St Andrews War Memorial Hospital, Spring Hill, Queensland, Australia\\
$^4$University of Queensland, School of Electrical Engineering, St Lucia, Queensland, Australia}

\ead{kristian.weegink@uqconnect.edu.au}
\begin{abstract}
In this paper we present an efficient model of the neuronal potentials recorded by a deep brain stimulation microelectrode (DBS MER)  in the subthalamic nucleus. It is shown that a computationally efficient filtered point process consisting of 10,000 neurons, including extracellular filtering closely matches recordings from 14 Parkinson's disease patients. The recordings were compared using their voltage amplitude distributions, power spectral density estimates and phase synchrony. It was found that interspike interval times modeled using a Weibull distribution with a shape parameter of 0.8, slightly non-Poisosnian, gave the best fit of the simulations to patient recordings. These results indicate that part of the `background activity' present in an DBS MER can be considered to be a very local field potential due to the surrounding neuronal activity.Therefore, the statistics of the interspike interval times modify the structure of the background activity.\\
\end{abstract}

\maketitle

\section{Introduction}

Deep brain stimulation (DBS) has become a common treatment for neurological movement disorders such as Parkinson's disease (PD)~\cite{Coyne06,McNames04}. DBS involves applying a pulsed electric field to a deep brain structure with an electrode positioned to within 1 mm accuracy. DBS applies chronic stimulation, using no feedback based on the patient's state, except for occasional clinical adjustment~\cite{Rouse11}. The subthalamic nucleus (STN) is a common target for the treatment of PD~\cite{Coyne06}.  The role of the STN as a DBS target for PD is not fully understood and for this reason many models of the brain have been developed to aid interpretation~\cite{Terman02,Rubin04,Feng07,Santaniello08}. Models of the STN may help further the understanding of DBS and in vivo neuronal activity in general. Improving the understanding of the role of the STN in DBS could benefit future stimulation therapies, including development of a closed-loop DBS strategy~\cite{Rouse11}. \\

To help determine whether the DBS stimulation electrode is implanted at the surgical target a micro-electrode recording (MER) is used~\cite{Coyne06,MERnoise}. A typical MER contains a baseline noise, we will call the ``background activity'', and spikes with a peak amplitude above the background activity~\cite{Garonzik04}. The DBS microelectrode has a tip diameter of $50~\mu m$. This allows the electrode to be used to both record and stimulate the target volume while preventing damage by minimizing current density. A consequence of this tip size is that it contains large background activity compared to a high impedance single neuron recording electrode ~\cite{SingleUnit}. Changes in the background activity can be used to confirm electrode placement~\cite{MERnoise}. This suggests that the ``noise'' component of a DBS MER has a neuronal component. We show numerically that a DBS MER can be considered a very local field potential (vLFP) with a superposition of electric fields at the tip from multiple neurons. The DBS MER background activity is due to ``competition'' between the filtering properties of the extracellular medium and the electrode geometry which leads to a large volume of neurons contributing.\\

Current STN models involving a large number of individual neurons are computationally intensive~\cite{Long10}. DBS MER models that simulate a single neuron with background noise are computationally efficcient but do not reflect neuronal noise processes. The aim of this paper is to develop a low-complexity neuronal model capable of modeling vLFPs recorded from the STN and thereby provide a better understanding of the origin of DBS MERs and the nature of the background activity. This model may also have use as an aid in confirmation of target volumes for DBS. In addition real-time adaptations of STN models are required for a feedback controller for DBS~\cite{Rouse11}.\\

The model we demonstrate in this paper is a filtered point process model of an STN MER. For the MER model 10,000 neurons are simulated and their electric fields are coupled to the micro-electrode with a spatial dependence. We demonstrate that this type of model is significantly more computationally efficient than current modeling techniques. To evaluate the accuracy of the model, we use the distribution of recorded amplitudes, linear fits of the modeled power spectrum to patient MER power spectra and comparisons of synchronous phase components.\\

This paper is structured as follows; section one provides the background and motivation for the modeling and analysis methods used. Section two outlines the methodology used for the simulations, the acquisition of patient recordings and the techniques used to quantify their comparison between them. Section three presents the results, section four includes a discussion of this work and finally section five provides the conclusions.

\subsection{Subthalamic Nucleus Models}

The STN is located in the basal ganglia. STN models vary from phase oscillators~\cite{Santaniello08} to spiking neuron models~\cite{Feng07}. Models that contain only a single spiking neuron and add random noise to produce an DBS MER~\cite{Santaniello08}. These models assume a spectrally ``white'' background noise. How neuronal activity changes this type of background activity hasn't previously been modeled. To overcome this a large number of individual spiking neurons are required in the simulation. The conductance based Hodgkin-Huxley (HH) model used in~\cite{Terman02} can be used to simulate individual STN neurons. This model is computationally intensive and requires the basal ganglia network to be modeled to produce correct spike timing~\cite{Terman02}.  This whole basal ganglia model is required to determine the STN firing times. To overcome this requirement of a large model we propose to use a filtered point process (FPP) model of the STN firing times.

\subsection{Filtered Point Process Models}
Stochastic models reduce an observable variable of a neuron, such as the timing of spikes, from being described by a deterministic equation, e.g. HH, to being drawn from a probability distribution, e.g. FPP. For features, like spike timing, the probability distribution used depends upon the network and noise inputs to a neuron. For the FPP each neuron is considered independently. The neuron action potential shape is used as the filter and occur at points in time described using only a probability distribution~\cite{Perkel67a,Perkel67b}.\\

A filtered renewal process is a special type of FPP where the time interval between two spikes (inter-spike interval, ISI) is drawn from a common distribution for each ISI~\cite{Bair95}. For a renewal FPP each ISI time is assumed independent of any previous times, that is to say the ISI times are independent identical distributions. For a filtered renewal process only the shape of the action potential and the probability distribution that describes the ISI is required to model a neuron~\cite{Banta64}. This significantly reduces the complexity as compared to other neuronal models such as~\cite{Feng07}.\\

The firing times for neurons is often described by a Poisson process, however there are many counter examples of non-Poisson neuron firing patterns~\cite{McKeegan02,Cateau06}. A Poisson process description cannot capture the cell behaviors seen in the STN such as bursting, uniform and periodic firing~\cite{IsraelSTN}. We propose that a  Weibull distribution is a more suitable ISI distribution for reproducing this range of STN behaviors.\\

\subsection{Electric Field Models for Extracellular Recordings}
The coupling of each neuron to the micro-electrode is dependent on the distance of that neuron to the electrode and the properties of the extracellular fluid in-between~\cite{Bedard09}.  As the electric field from the action potential propagates to the electrode from the neuron it passes through the extracellular space which has varying conductivity and permittivity. Finite element models (FEM) have been created to describe the electric field of neurons as it propagates through the extracellular medium~\cite{NeuronEAF}. The FEM simulations show that the extracellular medium causes filtering and attenuation of the potential measured at the electrode from different neurons. The spatial composition of the extracellular medium is also required to use this method and these models are too computationally expensive to implement for a efficient model. computational complexity can be reduced by assuming average properties of the extracellular medium~\cite{Bedard04}. This also removes the need to define the exact extracellular composition for each neuron-electrode interaction. The average extracellular filtering of the neuronal electric field at the electrode can be described by the complex transfer function ~\cite{Bedard09,Bedard04}. This assumes a point current source and a decoupled magnetic field. To obtain the transfer function is computationally expensive as a numerical integral needs to be calculated for each frequency component and for all neuronal positions. For an efficient model without large amounts of pre-calculated results an analytical form of the impedance is required. A circuit model simplification of this extracellular filtering can be used~\cite{Circuit2, Circuit1}. The effect of the radial distance to the electrode for each neuron is reduced to a ``seal'' resistance. This type of model also includes the frequency effects of the electrode geometry with Faradic resistance and capacitance. Therefore, we propose to use this neuron-electrode coupling model in this paper (described in section 2).

\subsection{Analysis of Recordings}
Several different analysis methods are used to compare the simulations and patient recordings. A standard method for analysing an MER is to use spike sorting~\cite{IsraelSTN}. Spike sorting removes background noise, so it is unsuitable for analysing the neuronal noise component of noise. To compare the simulations including noise to the {\it in vivo} recordings, statistically expected properties need to be used. Using renewal theory~\cite{Banta64} it has been shown that a filtered renewal process  has a  statistically expected power spectral density (PSD). The PSD can be written as a function dependent on the impulse filter (the action potential) and ISI probability density function. For this reason we propose to use the PSD to compare the simulated MERs to patient MERs.

Since the MER simulation will be modeled as a stochastic process in ISI times, the voltage history will not be deterministic. Random processes can be compared using statistical moments, like the mean and variance.  In this paper the KS test, \cite{KStest}, is used to compare the voltage distributions of simulated MERs to patient MERs.

Another feature of a stochastic process is random phase. To look at the phase properties of neural signals the component synchrony measure can be used \cite{CSM}.  This measure is also used in this paper.

\section{Methods}
The methods of this paper are organized into three sections. Section 2.1 contains patient information and the surgical method used to obtain the recordings. Section 2.2 describes how the simulations were performed. Section 2.3 contains the analysis methods used to compare simulations to patient recordings.

\subsection{Patient Recordings}

Nine participants (five male, four female) with idiopathic PD who were considered suitable for the implantation of bilateral permanent stimulators in the STN were included in this study. The patient age was $67\pm5$ years, with disease duration of $14\pm6$ years. Participants were all right handed and had no further neurological impairment. The participants had undergone psychiatric screening prior to DBS surgery. A summary of the patients is given in Appendix \ref{table:patients}.\\

The dorsolateral aspect of the STN was targeted using a Cosman-Roberts-Wells frame-based stereotactic frame with coordinates based on CT images fused with 3T MRI t1 and FLAIR sequences. The electrode placement was confirmed inter-operatively by an MER.  The surgical procedure is described in detail in~\cite{Coyne06}. Tungsten microTargeting$^{\textregistered}$ electrodes (model mTDWAR, FHC, Bowdoinham, ME) with a tip diameter of less than $50\mu m$ were used for the MER acquisition. The electrodes had a typical impedance of $0.5 (\pm 0.15) M\Omega$ at $1kHz$. A LeadPoint\texttrademark~  system (Medtronic Inc., Minneapolis,MN) was used to record the signals at a sampling rate of $24 kHz$. Three filters applied (high pass$:500Hz$ first order, low pass$:5kHz$ first  order  and anti-aliasing$:5KHz$ fourth order) as recommended by Medtronic. Each MER was recorded during resting phases, when the participant was lying still and not performing any cognitive or movement tasks.

\subsection{Simulations}

\begin{figure}[ht!]
  \centering
    \includegraphics[width=1\textwidth]{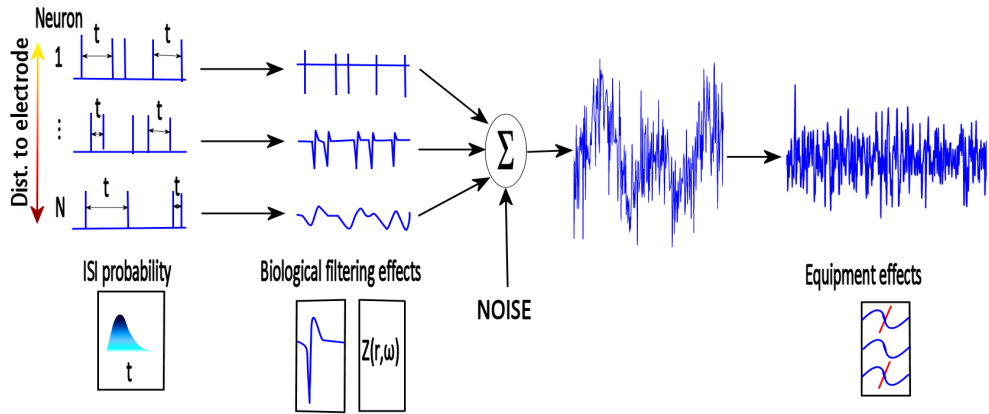}
  \caption{A summary of the method used for the simulations. Spike trains for each neuron are produced using the action potential shape and ISI probability distribution. These spike trains are filtered based on their distance from the electrode. The filtered spike trains are then summed together and noise is added before passing the signal through filters based on the equipment used in acquisition.}
\label{figure:Methodefig}
\end{figure}

The simulation method presented here are an extension of those presented in \cite{Kristian1}. A summary of the simulation method is given in Figure \ref{figure:Methodefig}. Simulations were performed for 10,000 neurons over one second to model the patient recordings. The use of 10,000 neurons is based on calculations in \cite{Kristian1}. However we propose to validate this numbe by calculating the RMS value of the MER simulations for different numbers of neuron.\\

All steps in the simulation were performed on a PC with a quad core 1.73 GHz processor and 8 GB or RAM using 64 bit MATLAB 7.14.0 (R2012a) \cite{MATLAB2010}. A time step of $1/24,000~s$ was used to match the sampling rate of the patient recordings.\\

\begin{figure}[ht!]
  \centering
    \includegraphics[width=0.7\textwidth]{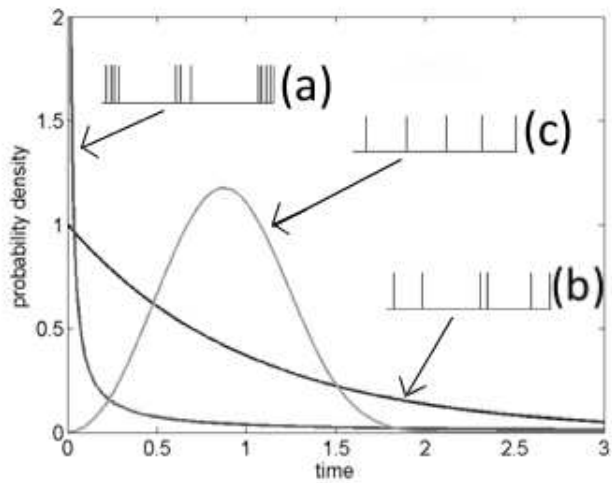}
  \caption{Examples of the Weibull distribution. The neuron can exhibit burst firing (a), Poisson statistics (b) or semi periodic firing(c) depending upon the shape parameter, $c$.}
\label{figure:wiebull}
\end{figure}

The STN behaviors is modeled by assuming the ISI times form a random variable drawn from a Weibull distribution in time:

\begin{equation}
	P(t) =
	\begin{cases}
		\left(\frac{t-t_r}{\lambda}\right)^{c-1}\frac{c}{\lambda}e^{\left(\frac{t-t_r}{\lambda}\right)^{c}}  & t>t_r, \\
		0 &\mbox{otherwise,}
	\end{cases}
	\label{eqn:Weibull}
\end{equation}

where $P(t)$ is the probability density function for an ISI and $\lambda$ is the scale parameter that controls the firing rate. The shape parameter $c$ controls the neuronal behavior; with $c<1$ generating burst firing, $c=1$ Poisson statistics and $c>2$ firing times with a common mode, as shown in figure \ref{figure:wiebull}. In the limit as $c\rightarrow\infty$ periodic behavior emerges. The parameter $t_r$ controls the refractory time of the neuron, preventing another action potential occurring within this period.\\

One second MER simulations were generated to match the length of patient recordings. The time series of Kronecker-delta pulses are first created by drawing the ISI times from the Weibull distribution shape parameter, \begin{it}c\end{it}, values of 0.5, 0.8, 1, 10 and 100. A refractory time of 5 ms was used \cite{IsraelSTN}.  Neuronal current time series are produced by convolving the Kronecker-delta pulses with the action potential shape. The time series was generated for each neuron independently. \\

The action potential shape was generated by numerically solving a Hodgkin and Huxley model using a variable order solver (ODE15s \cite{MATLAB2010}). The Hodgkin and Huxley model parameters used were for the medium spiny neuron based on~\cite{Terman02,Kristian1}: 

\begin{equation}
\begin{split}
C_m= & \frac{dV}{dt}=-g_L(V-v_L)-g_Kn^4(V-v_K)-g_{Na}m^3h(V-v_{Na}) \\ & -G_Ta^3b^2(V-v_{Ca})-g_{Ca}s^2(V-v_{ca}),
\end{split}
\label{eqn:HH}
\end{equation}

where $C_m$ is the membrane capacitance (1 pF/$\mu$m); $g_L$, $v_L$  are the leak conductance and reversal potential (2.25 nS/$m^2$ and -60.0 mV respectively); $g_K$, $v_K$  are the $K^+$ conductance and equilibrium potential (45 nS/$m^2$ and -80.0 mV respectively); $g_Na$, $v_Na$  are the $Na^+$ conductance and equilibrium potential (37.5 nS/$m^2$ and 55.0 mV respectively); $g_T$ is a low-threshold T-type $Ca^{2+}$  conductance (0.5 nS/$m^2$); and $g_Ca$, $v_ca$ are a high-threshold $Ca^{2+}$ conductance and a $Ca^{2+}$ equilibrium potential (0.5 nS/$m^2$ and 140.0 mV respectively). The gating variables $n$, $m$, $h$, $a$ and $b$ follow the differential equations and parameters given in~\cite{Terman02}. This produces the filter function used for each neuron.\\

\begin{figure}[ht!]
  \centering
    \includegraphics[width=0.7\textwidth]{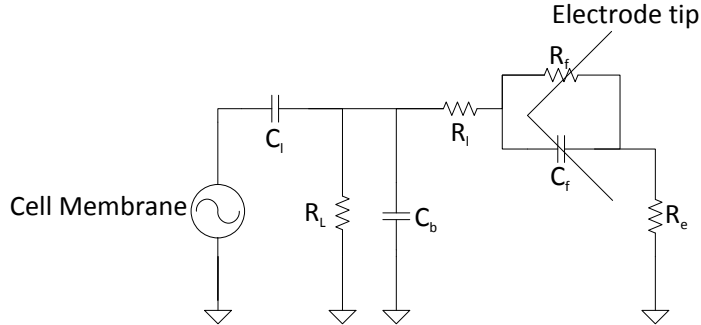}
  \caption{The circuit configuration used for coupling each neuron to the electrode.}
\label{figure:circuit}
\end{figure}

Each neuron is modeled as a point source, with the current being generated from the axon hillock. The current time series was then filtered using an impedance based on the distance of the neuron from the electrode to find the potential contributed by each neuron. 

\begin{equation}
V_\omega(r)=I_\omega Z_\omega(R),
\label{eqn:Ohms}
\end{equation}

where $I_\omega$ is the frequency component of the current at the neuron. The impedance filter $Z_\omega$ is found by determining the transfer function $I_\omega/V_\omega$ for the circuit model for the neuron-electrode interaction, shown in Figure \ref{figure:circuit}. This circuit model evaluates the propagation of the electric field through the extracellular medium, and is not an actual electron current. Circuit element values in this model depend on the radial distance between the electrode and neuron, the size of the electrode tip and the impedance of the electrode, where $C_l$ is the membrane-electrolyte interface capacitance, $R_L$ is the body resistance to ground (the spread of the field from the neuron), $C_b$ is the body's capacitance, $R_l$ is the resistance between the cell and the electrode (seal of the electric field by the neuron to the electrode \cite{Circuit2, Circuit1}), $R_f$ and $C_f$ are the electrode Faradic resistance and capacitance and $R_e$ is the electrode resistive load. The voltage for the recording, $V_\omega$, is taken across the load resistance.\\

The neuron radial distribution $N(R)$ is randomly generated using a uniform spatial distribution with density, $\rho=10^5 cm^{-3}$~\cite{IsraelSTN}.

\begin{equation}
N(R)=4\pi r^2\rho.
\label{eqn:dist}
\end{equation}

The complete time series from all neurons are then summed together linearly to create the potential across the electrode.\\

Thermal white noise (Johnson-Nyquist noise) is added to the electrode to match experimental conditions. The statistics of the noise are described by:

\begin{equation}
\left<V\right>=0, \qquad \left<V^2\right>=4k_BTR(f)\Delta f,
\label{eqn:vartherm}
\end{equation}

where $k_B$ is Boltzmann's constant, $T$ is the temperature (assumed to be average body temperature 37$^o$C), $R(f)$ is the electrode resistance, $\Delta f$ is the bandwidth and $\left<\cdots\right>$ represents the time average. The product $R(f)\Delta f$ is calculated by integrating the product of $R(f)$ with the gain function $G(f)$ of the equipment over frequency:
\begin{equation}
 R(f)\Delta f=\int_0^{\infty}{R(f)G(f) df}.
 \end{equation}

To match simulations to the surgical conditions the simulated voltage time series is passed through three filters described in section 2.1. The filters are models  of the two software filters with a 500Hz first order high pass, 5kHz first order low pass and and the hardware 5kHz fourth order anti-aliasing filter.

\subsection{Comparative analysis of modeled and patient recordings}

The analysis of the model is broken into three sections. The sections look at the distribution of recorded amplitudes in the time domain, linear fits of the modeled PSD estimate to patient PSD estimates and comparisons of synchronous phase components. For the analysis the tests were averaged over multiple recordings from the same patient and the two separate patient sides were analyzed separately. Patient recordings that contained movement artifacts, defined by amplitude $>~10~mV$, or had recording times less than 1~s were removed from the analysis. After this removal process, 84 MERs from 14 patient-hemispheres were analyzed.

\subsubsection{Test of Voltage Distributions:}

The first test performed was a two sided Kolmogorov-Smirnov (KS) test on the distribution of the voltages over time. This test was used to check if the voltage amplitude distribution for the simulations have different distributions to the voltage amplitude distribution seen in a patient recordings. The KS test produces a p-value for the null-hypothesis that the amplitudes are drawn from the same distribution.

\subsubsection{Power Spectrum Comparisons:}

PSDs for the recordings and simulations were calculated using the Welch method with a Hamming window~\cite{MATLAB2010}. The PSDs obtained using the five different simulation parameters were compared to the 14 patient hemisphere recordings using linear regression. The linear regression used the value of the patient PSD against the simulated PSD for each frequency. The correlation coefficient was used to asses the goodness of fit.

\subsubsection{Phase comparisons:}

The individual recordings are divided into 100 ms non-overlapping sections. The variance of the phase for each frequency for each section was found using~\cite{CSM}:
\begin{equation}
\mathrm{var}\lbrace\phi(m)\rbrace= 1 -\left[\frac{1}{N}\sum_{i=1}^N{\cos{\phi_i(m)}}\right]^2-\left[\frac{1}{N}\sum_{i=1}^N{\sin{\phi_i(m)}}\right]^2
\label{eqn:phasevar}
\end{equation}
The component synchrony measure (CSM) was then be calculated by using:
\begin{equation}
\mathrm{CSM}(m)=1-\mathrm{var}\lbrace\phi(m)\rbrace.
\label{eqn:CSM}
\end{equation}

\section{Results}

The results are divided as follows; Section 3.1 contains a summary of the computation time and features of MER simulation. Section 3.2 contains the results from the comparison with patient recordings.

\subsection{Simulations}

\begin{figure}[ht!]
  \centering
    \includegraphics[width=0.7\textwidth]{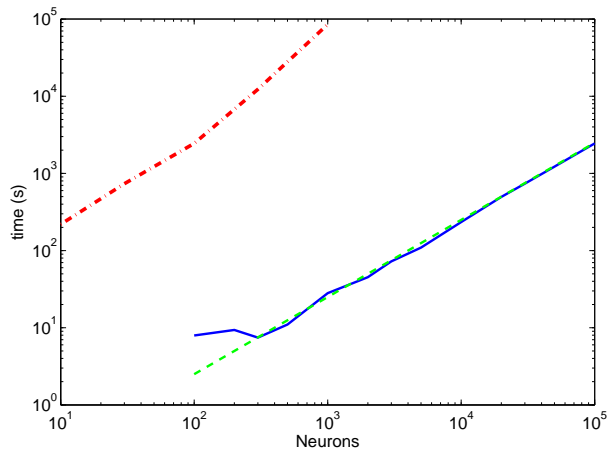}
  \caption{The computational time to simulate an MER using the method presented in this paper (solid), a Hodgkin and Huxley neural network (dot dash). A line with slope one (dashed) is layered on top to indication O(N).}
\label{figure:comptime}
\end{figure}

To illustrate the speed advantages of the FPP model over a deterministic HH model, a comparison of the time required to compute an MER using the proposed model and a coupled HH network is shown in figure \ref{figure:comptime}. The points are averaged over three runs. The dashed line is a line of unity slope, to show that the computational order of the FPP is approximately O(N). These simulations were performed on the same computer with no other tasks running. Due to the long computation time, no simulations of the Hodgkin and Huxley network with over 1,000 neurons were performed. The comparison of the computational time compared to neuron number shows that the FPP model is significantly faster than the equivalent Hodgkin and Huxley network mode.\\

\begin{figure}[ht!]
  \centering
    \includegraphics[width=0.7\textwidth]{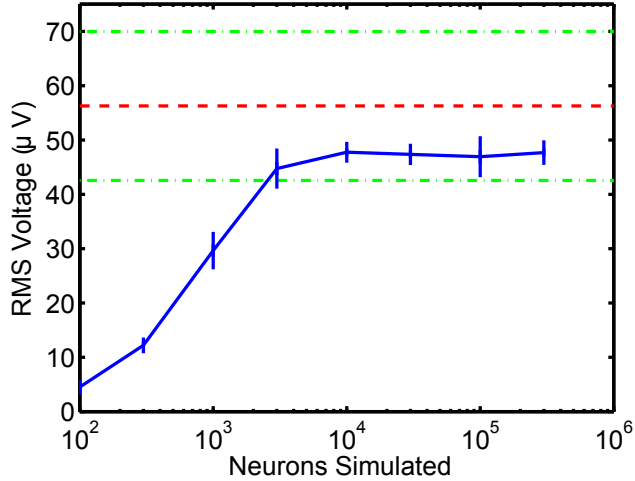}
  \caption{The effect of changing the number of neurons simulated on the RMS value of the MER. The dashed lines represent the mean RMS (dark line) of patient recordings and one standard deviation (light line).}
\label{figure:NeuralNoise}
\end{figure}

Figure \ref{figure:NeuralNoise} shows how the RMS value of the simulated MERs changes as the neuron number is changed. Above 3,000 neurons the RMS value begins to plateau. The peak RMS value approaches $49 \mu V$. This is within one standard deviation of the patient recordings mean RMS value of $56 \pm 12 \mu V$.\\

For the same recording length c=0.5 (small dotted line) shows the least power density across the band of interest (unfiltered region, $500~\textrm{Hz}<\omega<5000~\textrm{Hz}$) from the patient recordings. The other extreme of $c=100$ (dashed line) shows harmonic spikes. For $c=0.8$ the PSD has a more spread out frequency distribution compared to the other simulations. The PSD for $c=1$ follows the action potential power spectrum as expected due to Carson's theorem for a Poisson process. Although $c=0.8$ and $c=1$ have a very similar shape of their ISI times distribution (exponential), they display different distributions of power in their PSD estimates. PSD estimates for simulations using different shape parameters are shown in Appendix \ref{figure:SIMPSD}.

\subsection{Comparative analysis of modeled and patient recordings}

\begin{figure}[ht!]
  		\centering
    		\includegraphics[width=0.7\textwidth]{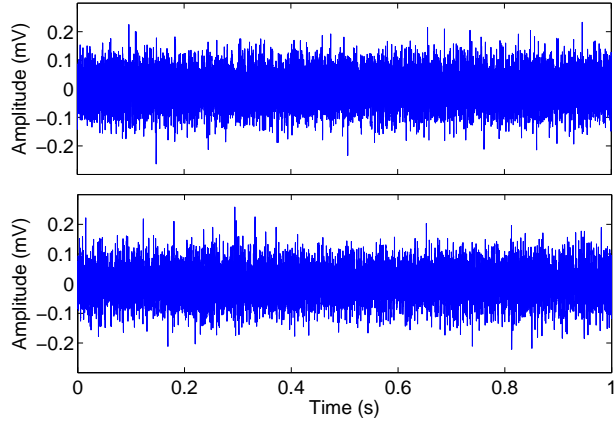}
  		\caption{Examples of a patient recording (top) and a simulated recording (bottom) on a coarse time scale with a simulation parameter $c=0.8$.}
		\label{figure:2MER}
\end{figure}

\begin{figure}[ht!]
  		\centering
    		\includegraphics[width=0.7\textwidth]{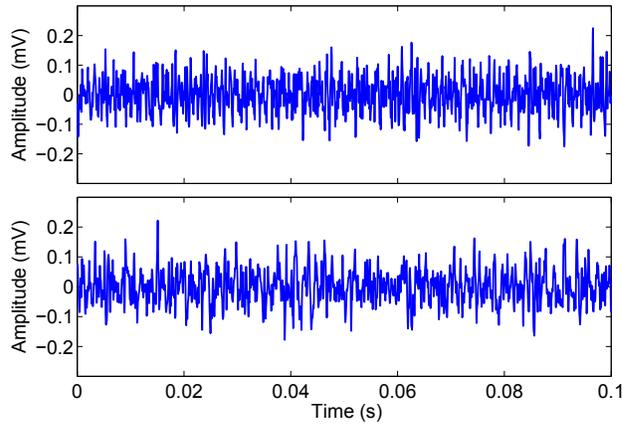}
  		\caption{Examples of a patient recording (top) and a simulated recording (bottom) on a fine time scale with a simulation parameter $c=0.8$.}
		\label{figure:2MERclose}
\end{figure}

Figure \ref{figure:2MER} and Figure \ref{figure:2MERclose} show comparisons of the patient DBS MER to a simulation with a Weibull shape parameter of 0.8, at two different time scales. These two comparisons visually show a similarity that scales (look similar) over the 1~s to 100~ms timescales.

\subsubsection{Test of Voltage Distributions:}

\begin{figure}[ht!]
  \centering
    \includegraphics[width=0.7\textwidth]{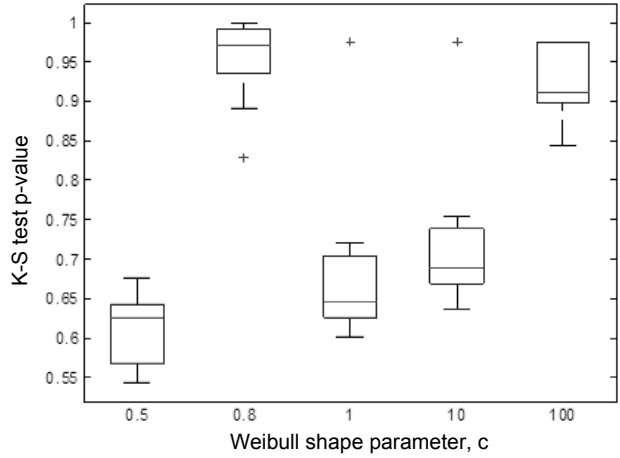}
  \caption{Box plot of the KS test p-value of each patient voltage distribution matching the simulation distribution for each shape parameter. The Box represents the 25 and 75 percentiles, the lines represent the maximum and minimum values, the midline represents the median value and the `$+$' represents outliers.}
\label{figure:KSBox}
\end{figure}

Figure \ref{figure:KSBox} shows a box-plot of p-values from the KS test on the voltage distributions of the 14 hemisphere recordings against simulation parameters $c=0.5$ to $c=100$.  As all p-values are above 0.1 we can not reject the null hypothesis that the variables are drawn from the same distribution. The highest median of p-values was for $c=0.8$, indicating that this empirical cumulative distribution function for the model has the smallest distance to the patient's empirical cumulative distribution function as measured by the KS test.

\subsubsection{Power Spectrum Comparisons:}

\begin{figure}[ht!]
  \centering
    \includegraphics[width=0.7\textwidth]{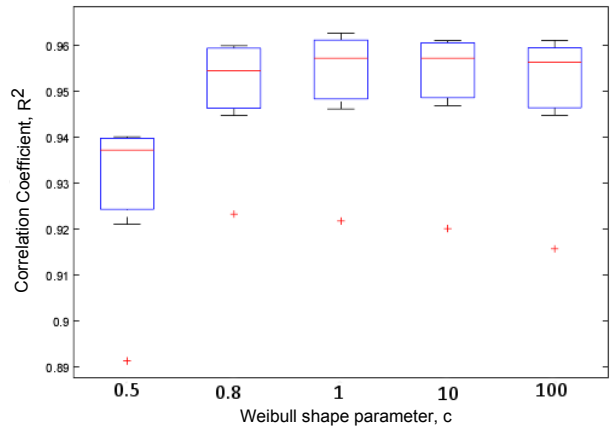}
  \caption{Box plot of the $R^2$ value from fitting each patient spectrum to the simulated spectrums for different c values.}
\label{figure:LinearBoxplot}
\end{figure}

 Linear regression of the simulated PSD against the patient MER PSD was used to asses the model fit to the patient recordings. Figure \ref{figure:LinearBoxplot} shows a box plot of the correlation coefficient for the linear fit for the 14 patient hemisphere recordings. The outlier point is patient 61 right side for all values of c. This figure also shows that the $R^2$ value is greater than 0.89 for all values of the shape parameter. The worst case fit has a shape parameter $c=0.5$. The model fits the patient recordings for $c\ge0.8$ to similar accuracy with $c=1$ having the highest value.

\subsubsection{Phase Comparisons:}

\begin{figure}[ht!]
  \centering
    \includegraphics[width=0.7\textwidth]{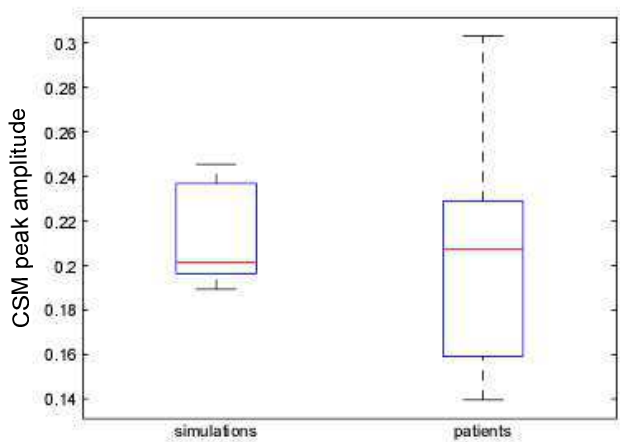}
  \caption{The amplitude of the highest peak from each CSM spectrum.}
\label{figure:CSMboxamp}
\end{figure}

The method used to see if there are any features in the phase spectrum of the recordings was the component synchrony measure (CSM). The amplitude of the largest peak in each CSM spectrum is shown in Figure \ref{figure:CSMboxamp} for both simulation and patient data. Figure \ref{figure:CSMboxamp} shows there are no peaks with amplitude above 0.3 in any of the recordings or simulations and therefore no significant phase structure.  

\section{Discussion}

The implications from the model presented in this paper are given in section 4.1. Section 4.2 looks at the results of the analysis and section 4.3 discusses the assumptions and limitations for the model with possible future extensions. Section 4.4 summarizes the discussion.

\subsection{Modeling}

The computational time of the FPP model diverges from O(N) at low neuron number seen in Figure \ref{figure:comptime}. This divergence from O(N) is due to the minimum time to initialize the simulation.\\

Other dynamic models of neurons, which reduce the complexity of the differential equations of the Hodgkin and Huxley model were not used to compare the computational time to this model. Computationally efficient spiking neuron (leaky integrate and fire and Izhekivich) models cannot produce accurate enough action potential shapes and are generally used to produce the correct spike timing~\cite{Izhikevich07a}. Because the PSD in the frequency range of interest has a contribution from the shape of the action potential these models were not considered.\\

The model proposed in this paper only produces the timing and shape of action potentials. The model does not account for any of the electrical activity below threshold that activates the spike. This type of activity, called sub-threshold oscillations, are typically low frequency (1-100Hz).  Slow oscillations are not clearly seen in the patient recordings due to the shape of the electrode ($50 \mu m$ tip) and the high pass filter at 500Hz. Due to these factors, subthreshold oscillations do not require modeling to accurately produce a DBS MER.\\

The change in RMS values in Figure \ref{figure:NeuralNoise} numerically shows that when choosing a neuron number over $\approx 10,000$ the extra neurons do not contribute significantly. Additional neurons above 10,000 do not contribute due the extracellular filtering effects, see Appendix \ref{figure:EAfilter}. Increasing the number of neurons places them further from the electrode. When the distance becomes too large their electric fields do not contribute to the recordings. 

\subsection{Analysis}

Three different methods were used to analyze the MER data. These methods are used to build a comprehensive comparison between the patient and simulated recordings. The first order analysis, using the voltage distribution, demonstrates matching behavior of the voltage over time. The second order analysis, using the PSD, allows the correlation properties of the model and patient recordings to be analyzed using different interspike interval statistics. The phase properties are used to verify the random phase assumption of a stochastic process.

\subsubsection{Voltage Distribution:}

\begin{figure}[ht!]
  \centering
    \includegraphics[width=0.7\textwidth]{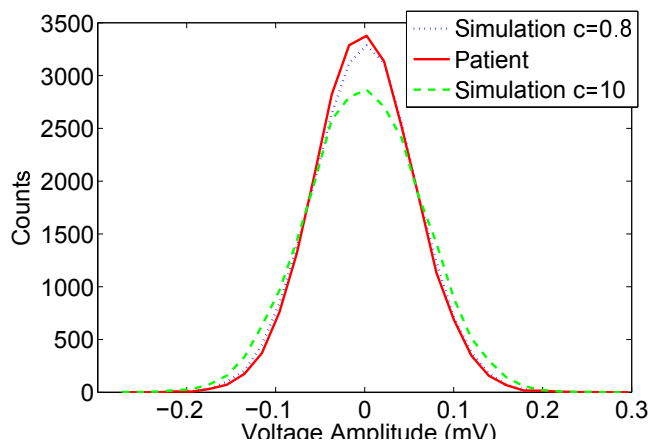}
  \caption{Examples of the voltage distribution for simulation c=0.8 (dotted), patient 32L (solid) and simulation c=10 (dashed)}
\label{figure:KStest}
\end{figure}

The KS test estimated the p-value for the null-hypothesis that voltage for the simulation and patient recordings are drawn from the same distribution. For all values of $c$ the test statistic, $p$, was greater then 0.1. This means that for each $c$ value on these ser of data we can not reject the null-hypothesis. However, the size of the p-value can then be used to indicate how close the distributions are to being the same. Examples of this is shown in Figure \ref{figure:KStest}. The KS test produces a p-value above $0.1$ meaning we cannot reject the null-hypothesis for both of these $c$ values. However, it can be seen that for $c=0.8$ (p=$0.98$) has a better correspondence tp patient recordings than $c=10$ ($p=0.72$). These results suggest that simulations with $c=0.8$ best match the patient recordings.  

\subsubsection{Power Spectral Comparisons:}

\begin{figure}[ht!]
  \centering
    \includegraphics[width=0.7\textwidth]{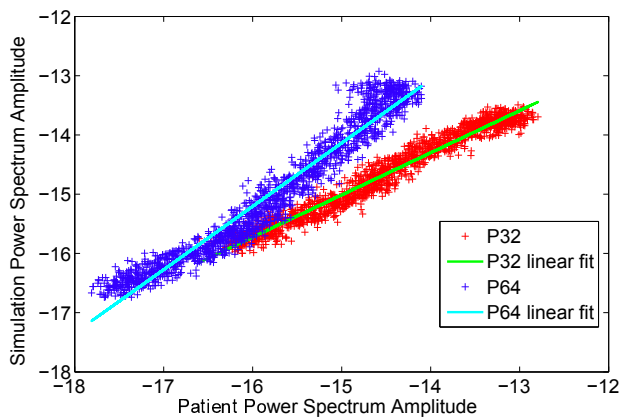}
  \caption{Example of linear fits of the patient frequency power versus simulation frequency power for P32L (light) and P61R (dark).}
\label{figure:linearregression}
\end{figure}

The linear regression of patient and simulated PSD, with Weibull shape factor, $c$, ranges from 0.8 to 100 and gives good agreement with the patient recordings. Assuming constant action potential shape between patients, the changes in inter-patient PSD estimates are indicative of changes in the ISI statistics. Recordings 32 left and 61 right had the best and worst fit to simulations with $c=0.8$, shown in Figure \ref{figure:linearregression}. Appendix \ref{figure:patientpsd} shows the patient PSD estimates with the best and worst fit and a PSD estimate from a simulation using $c=0.8$. The 95\% confidence interval is also plotted for five repeated recordings from the same patient. This visually demonstrates the accuracy of the model, as well as the variation in patient recordings. For comparison, Appendix \ref{figure:WnRegress} shows the regression of a patient PSD against white noise with equipment filtering effects. This regression has a low correlation coefficient, $R^2=0.0306$, indicating that the background of the patient recordings contain contributions from neural activity.\\

Variation in fits between the different patients data sets can be explained by the fact that an ``average'' electrode impedance of $0.5~M\Omega$ at 1kHz was used for the model. In reality the impedance changes slightly for each patient~\cite{Leadpoint}. This model could be used to improve the fit to individual patients by measuring the electrode impedance prior to recording.

\subsubsection{Phase Properties:}

Since a stochastic process in time will have random phase, the phase information should show no synchrony between any frequency components. CSM values above 0.5 show a significant amount of phase synchrony across the recordings at a specific frequency~\cite{CSM}. Figure \ref{figure:CSMboxamp} shows there are no peaks above 0.3 in the CSM spectra for the patients or simulations with patients being slightly more variable. This indicates that there is no phase synchrony present in either the model or patient recordings. \\

The difference in the distribution of CSM peaks in the model can be explained by an additional white noise source that isn't modeled. A white noise source would not affect the spectral color, as it would add power across all frequencies, however it can add noise to the phase spectrum that has a variance that scales with the amplitude of the source~\cite{CSM}. This could increase the CSM peaks that are not above the significance of 0.5 and would be spread across a wider frequency range than the simulation currently displayed.\\

\subsection{Limitations and Future Work}

The simplification of the modeled system improves the computational efficiency in simulating an MER but decreases the realistic features seen, e.g.~\cite{Izhikevich07a}. One feature that can be seen in neurons that this model fails to reproduce is synchronization in the neural network. Although several network features can be modeled through the shape of the ISI probability, there is no ability for two or more neurons to begin firing synchronously or to have any firing correlations.

Neurons can demonstrate a behavior called rate dependent action potential shape. This is where the shape of the action potential (both amplitude and frequency components) can change with the rate that the neurons fire~\cite{RateDependant}. If the target neuron displays rate dependent action potential shape, the model cannot account for this effect. Neurons that do display this behavior usually have two distinctive action potential types. One shape when the neuron is firing slowly and a sharp change to another other when the neuron is firing near its maximum rate \cite{RateDependant}. Assuming the neuron only fires in a particular rate range, the effect of rate dependent action potential shapes can be minimized by approximating a single waveform over that range.\\

\subsection{Summary}
For an MER acquired in surgery there is `competition' between the low pass frequency characteristics of the cellular environment and the high pass frequency characteristics of the electrode tip. This means that for an MER it is not only a single neuron in the immediate vicinity that contributes, like a single neuron recording~\cite{SingleUnit}, nor is it the low frequency behavior from the local structure of neurons, such as a local field potential~\cite{LFP}. Rather it is a vLFP as we have demonstrated.

Synchrony may be present between neurons, even though both the simulations and patient recordings do not indicate any phase synchrony. The effect of synchrony between the neurons would manifest as larger spikes in the time series of the MER. This type of behavior is not present in the model due to the neurons being modeled as independent. Future work will include neuron synchrony by including synchronous firing events as a second process with different statistics to the individual neurons ISI.

\section{Conclusion}

In this paper we have proposed an efficient model of an MER acquired from the STN during DBS implantation for PD. We have shown, on a set of 40 recordings from 14 patient hemispheres, that this computationally efficient, 10,000 neuron model fits recordings from patients well in terms of the voltage amplitude distribution, the power spectral estimates and phase synchrony. These results indicate that part of the ``noise'' present in an MER can be considered a vLFP and is due to the neuronal activity surrounding the electrode. This noise was shown to be dependent on a model parameter that controlled the ``shape'' of the ISI time distribution. When the ISI is modeled using a Weibull distribution the shape parameter that best fit the data across all the tests was $c=0.8$, slightly non-Poissonian. This indicates that rather than following a Poisson process, the neurons are firing with inter-spike intervals that more closely follows a stretched exponential distribution.

\section*{Acknowledgments}

The authors are greatly indebted to PD specialists of St. Andrew's War Memorial and The Wesley Hospitals, Australia for their motivation, guidance and interdisciplinary expertise. Research funds were provided by The University of Queensland and Medtronics. A. P. B. is the recipient of an Australian Research Council Future Fellowship (FT110100623). K. J. W. is the recipient of an Australian Postgraduate Award.

\clearpage
\appendix
\section{Patient Summary}
\begin {table}[ht!]
\caption {Summary of participants for whom MER recordings were used for the validation of the model.} \label{tab:patients} 
\begin{footnotesize}
\begin{center}
\begin{tabular}{|l|l|l|l|l|l|l|l|l|} \hline
\label{table:patients} Partic-&Age&Gender&Edu-&Handed-&Disease&Severity&UPDRS III&Side of\\
ipant& & &cation &ness &duration&(H\&Y)&on Score&MER\\ \hline\hline
32 &73 &M &14 &R &8 &NA &NA &Left \\ \hline
38 &58 &M &11 &R &11 &2 &3 &Bilateral \\ \hline
53 &71 &M &13 &R &16 &4 &20 &Bilateral \\ \hline
61 &71 &F &10 &R &17 &3 &17 &Bilateral \\ \hline
69 &66 &F &14 &R &22 &NA &NA &Bilateral \\ \hline
74 &65 &F &7 &R &15 &2 &11 &Bilateral \\ \hline
103 &62 &M &9 &R &20 &2 &8 &Right \\ \hline
104 &71 &M &10 &R &3 &NA &NA &Bilateral \\ \hline
\end{tabular}\\
\end{center}
Note: H\&Y = Hoehn and Yahr; UPDRS = Unified Parkinson's Disease Rating Scale; ``on'' refers to on stimulation and on medication; R = right; NA not available.\end{footnotesize}\end{table}
\section{Results}

\begin{figure}[ht!]
  \centering
    \includegraphics[width=0.7\textwidth]{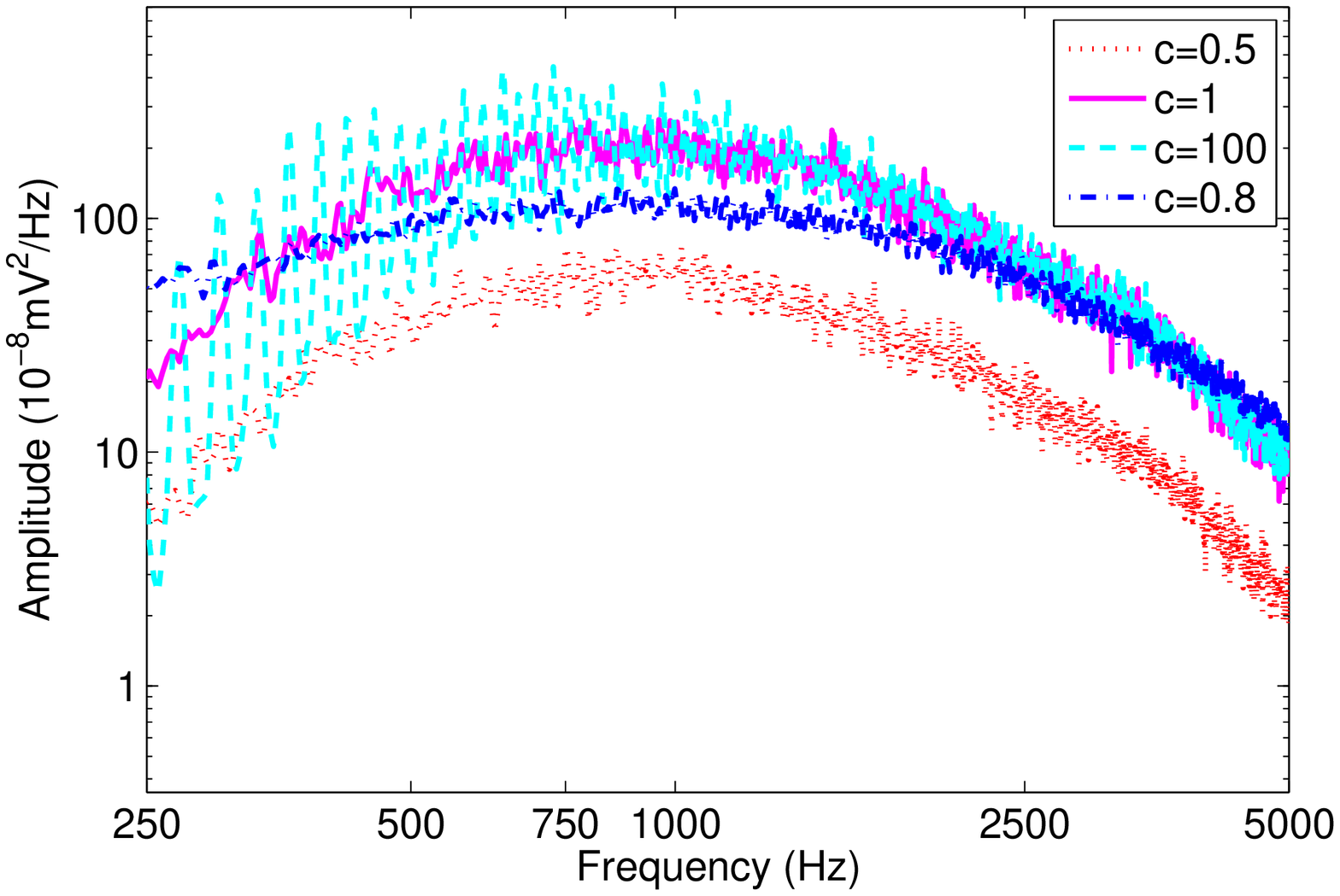}
  \caption{PSD estimates for the simulations using $c=0.5$(dots), $c=1$(solid),$c=100$(dashed) and $c=0.8$ (dot dash). }
\label{figure:SIMPSD}
\end{figure}

\begin{figure}[ht!]
  \centering
    \includegraphics[width=0.7\textwidth]{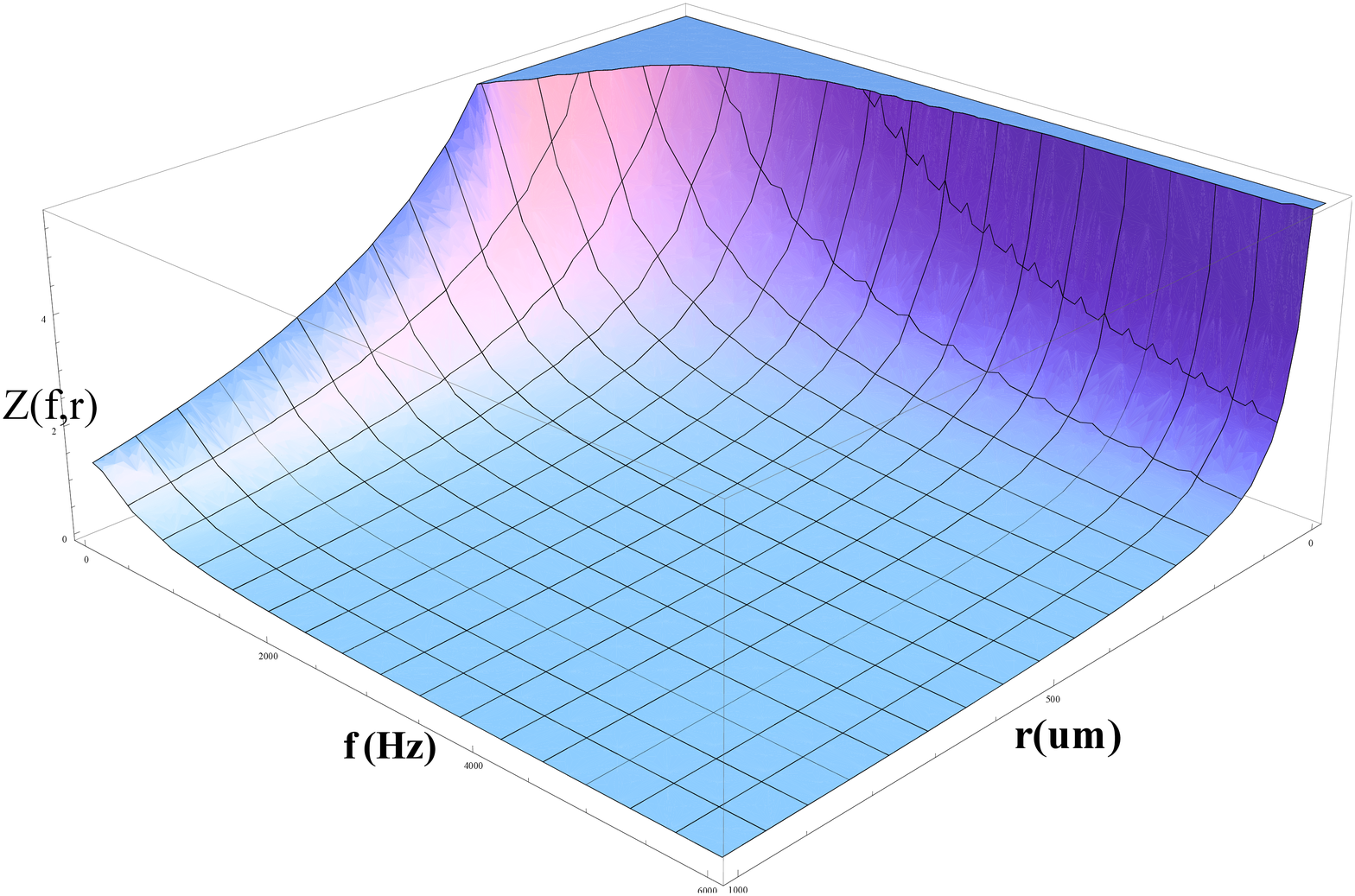}
  \caption{Plot of the extracellular transfer function from the circuit model in Figure \ref{figure:circuit}}
\label{figure:EAfilter}
\end{figure}

\begin{figure}[ht!]
  \centering
    \includegraphics[width=0.7\textwidth]{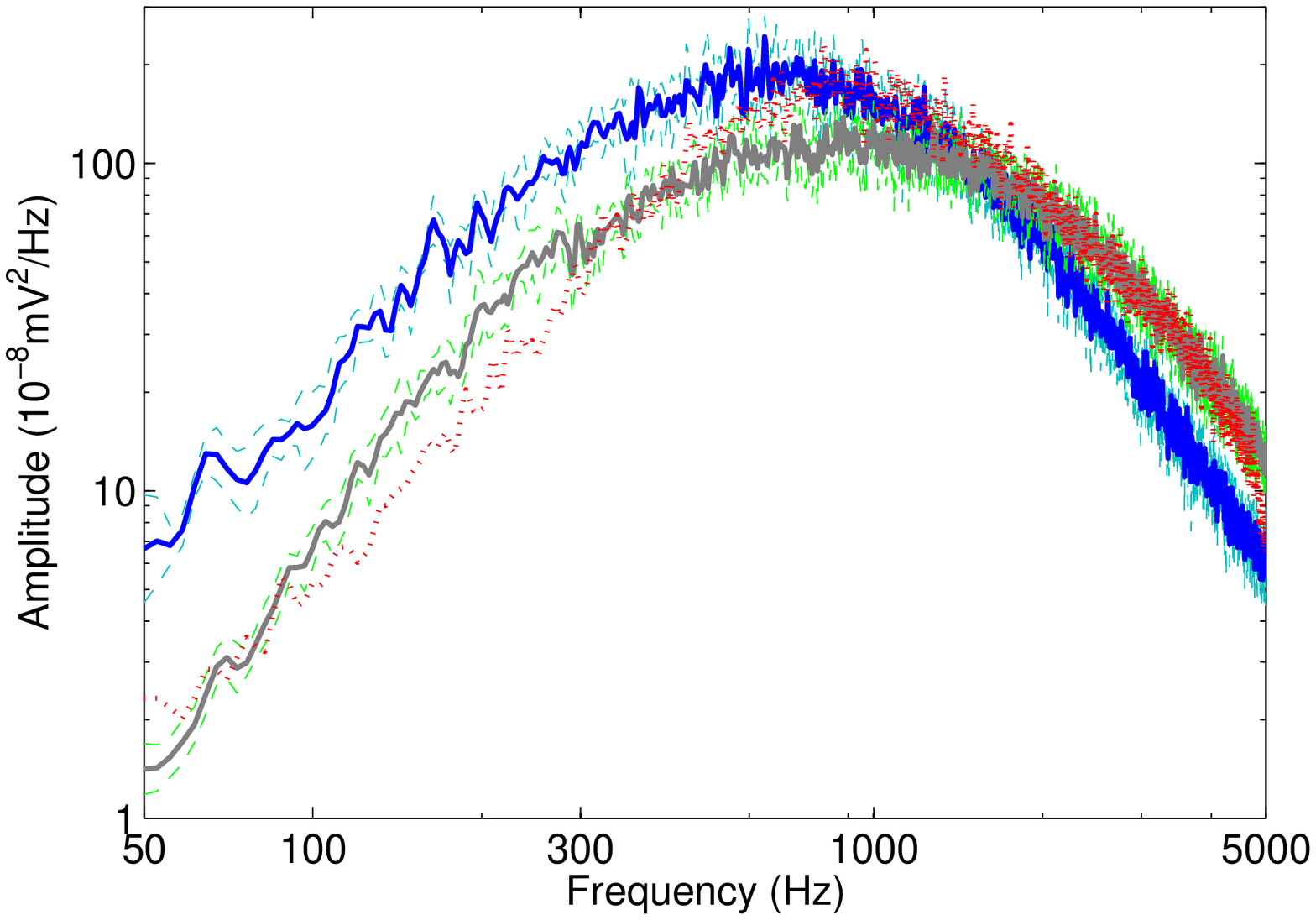}
  \caption{Examples of the spectral estimates from two patient recordings (P32L light, P61R dark) with 95$\%$ confidence interval (dashed) and a simulation with $c=0.8$ (dots).}
\label{figure:patientpsd}
\end{figure}

\begin{figure}[ht!]
  \centering
    \includegraphics[width=0.7\textwidth]{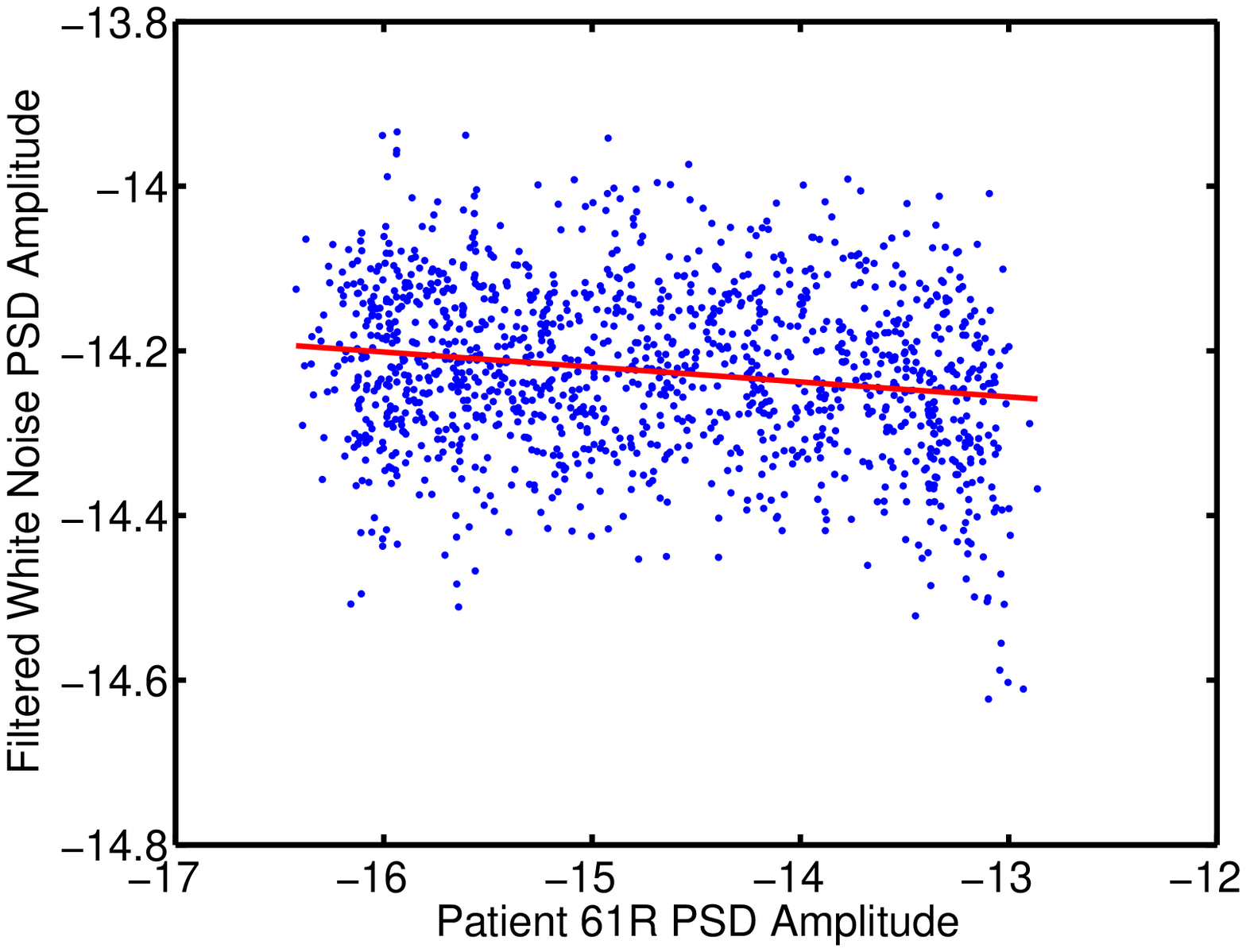}
  \caption{Linear regression of the PSD of the P61R with white noise passed through the equipment filters. The regression line has the form $y=-0.0182x-14.4920$ with a correlation coefficient $R^2=0.0306$.}
\label{figure:WnRegress}
\end{figure}
\clearpage
\section*{References}{}

\bibliography{PPpaper}
\bibliographystyle{plain}

\end{document}